\documentclass{article}

\PassOptionsToPackage{numbers, sort&compress}{natbib}
\usepackage[preprint]{neurips_data_2022}

\usepackage[utf8]{inputenc} 
\usepackage[T1]{fontenc}    
\usepackage{hyperref}       
\usepackage{url}            
\usepackage{booktabs}       
\usepackage{amsfonts}       
\usepackage{nicefrac}       
\usepackage{microtype}      
\usepackage{xcolor}         
\usepackage{graphicx}       
\usepackage{caption,stackengine}
\usepackage{wrapfig}

\title{Evaluating histopathology transfer learning with ChampKit}

\author{
  Jakub R.~Kaczmarzyk \\
    Department of Biomedical Informatics \\
    Stony Brook Medicine \\
    \texttt{jakub.kaczmarzyk@stonybrookmedicine.edu} \\
    \And
    Tahsin M. Kurc \\
    Department of Biomedical Informatics \\
    Stony Brook Medicine \\
    \texttt{tkurc@stonybrookmedicine.edu} \\
    \And
    Shahira Abousamra \\
    Department of Computer Science \\
    Stony Brook University \\
    \texttt{sabousamra@cs.stonybrook.edu} \\
    \And
    Rajarsi Gupta \\
    Department of Biomedical Informatics \\
    Stony Brook Medicine \\
    \texttt{rajarsi.gupta@stonybrookmedicine.edu} \\
    \And
    Joel H.~Saltz \thanks{Co-supervisors.} \\
    Department of Biomedical Informatics \\
    Stony Brook Medicine \\
    \texttt{joel.saltz@stonybrookmedicine.edu} \\
    \And
    Peter K.~Koo \footnotemark[1]  \\
    Simons Center for Quantitative Biology \\
    Cold Spring Harbor Laboratory \\
    \texttt{koo@cshl.edu}
}

\begin{document}

\maketitle

\begin{abstract}
Histopathology remains the gold standard for diagnosis of various cancers. Recent advances in computer vision, specifically deep learning, have facilitated the analysis of histopathology images for various tasks, including immune cell detection and microsatellite instability classification. The state-of-the-art for each task often employs base architectures that have been pretrained for image classification on ImageNet. The standard approach to develop classifiers in histopathology tends to  focus narrowly on optimizing models for a single task, not considering the aspects of modeling innovations that improve generalization across tasks.
Here we present \textit{ChampKit} (Comprehensive Histopathology Assessment of Model Predictions toolKit): an extensible, fully reproducible benchmarking toolkit that consists of a broad collection of patch-level image classification tasks across different cancers. ChampKit enables a way to systematically document the performance impact of proposed improvements in models and methodology. ChampKit source code and data are freely accessible at \url{https://github.com/kaczmarj/champkit}.
\end{abstract}

\section{Introduction}

\begin{figure}[ht]
  \centering
  \includegraphics[width=5.3in]{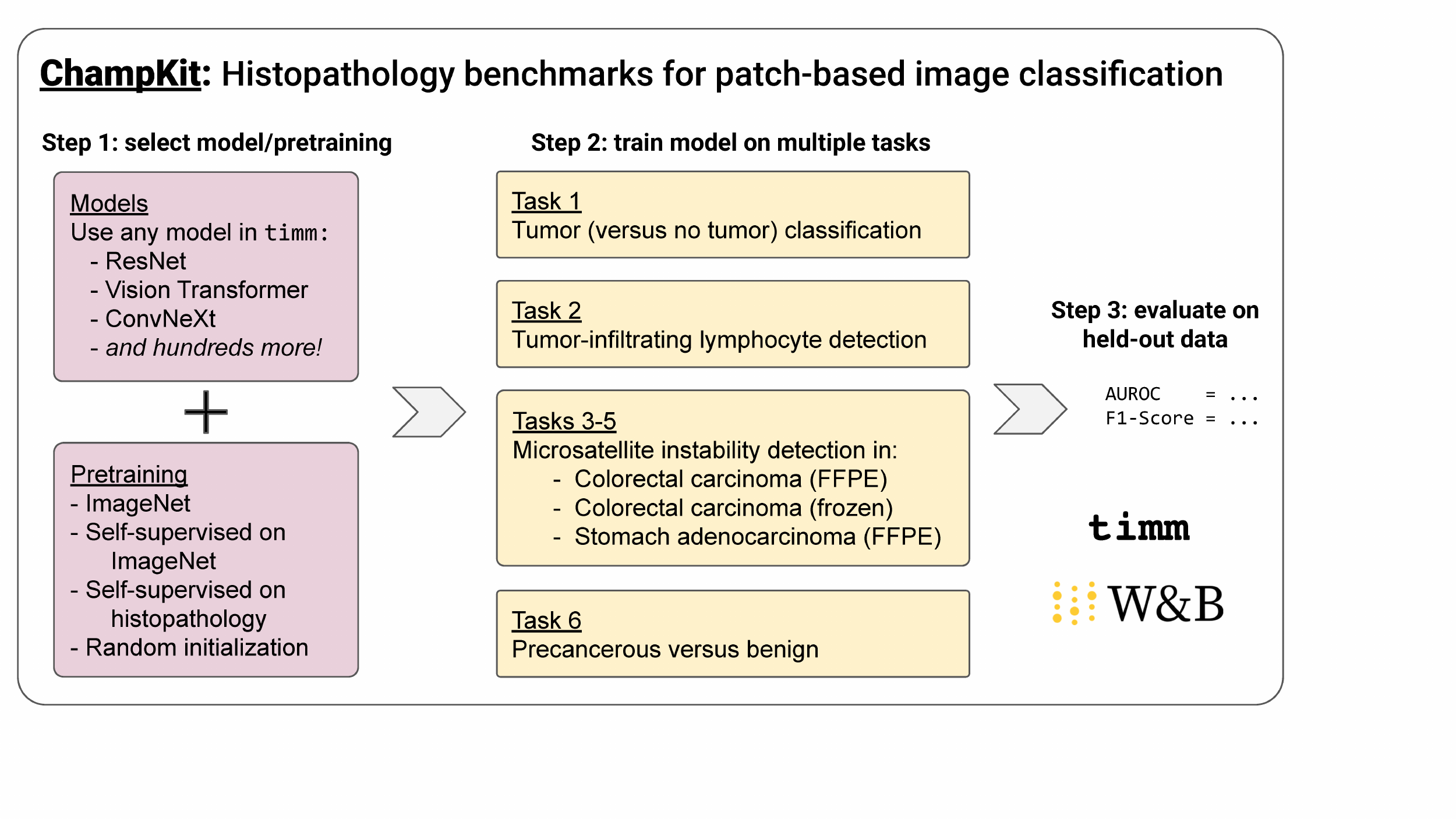}
  \caption{Overview of ChampKit. ChampKit enables systematic comparisons of model architectures and transfer learning across several patch-based image classification datasets. First, users select a model and pretrained weights from those available in timm \citep{rw2019timm} or a custom model with specfication of pretrained weights or random initialization. Second, the models are trained on multiple tasks using identical training hyperparameters. Third, the trained models are evaluated on held-out test data for each task. Performance is tracked with Weights and Biases (W\&B) \citep{wandb}.}
  \label{champkit-figure}
\end{figure}

Histopathology is the gold standard for diagnosing many types of cancers. There have been many efforts to apply computer vision techniques to histopathology throughout the years and it has gained more popularity with the emergence of deep learning \citep{van2021deep, banerji2022deep, sultan2020use, jimenez2017analysis, srinidhi2021deep, deng2020deep, hamidinekoo2018deep, coudray2018classification}. There are many biologically and clinically important features that deep learning has been used to predict, including presence of invasive tumors \citep{liu2017detecting,  wang2016deep, lee2018robust, awan2018context, iizuka2020deep, kwok2018multiclass}, presence of tumor infiltrating lymphocytes \citep{saltz2018spatial, abousamra2022deep, lu2020deep, linder2019deep, meirelles2022effective, baid2022federated, amgad2020report}, and prediction of microsatellite instability \citep{kather2019deep, echle2020clinical, muti2021development, echle2021deep, cao2020development}.

Whole-slide histopathology images are large (e.g., 100,000 by 100,000 pixels) relative to images used in standard deep learning applications like ImageNet \citep{ILSVRC15}. In practice, they are processed into smaller patches for use in machine learning. While most data processing and modeling happens at the patch-level, the overall prediction task falls either into patch-level classification or whole-slide (i.e. patient-level) classification. Whole-slide classification is used to predict a single label based on multiple image patches (e.g., response to cancer treatment, time to death, presence of somatic mutation). By contrast, patch-level classification aims to classify local features within individual patches, each of which has a label (e.g., detection of prognostic biomarkers and cancer cells). Whole-slide classification requires a different modeling approach, such as multiple-instance learning, whereas patch-level classification extends more naturally to traditional computer vision models, where each image is treated independently. Here we focus on patch-level image classification tasks.

Deep learning models for patch-level classification tasks tend to employ pretraining on ImageNet \citep{abousamra2022deep, kather2019deep, sharmay2021histotransfer, springenberg2022cnns, kassani2019classification, saltz2018spatial}, despite the fact that the original features are natural images. Recently, there has been growing interest in leveraging the success of self-supervised pretraining to directly learn features from unlabeled histopathology datasets \citep{ciga2022self, wang2021transpath}. Many claims of performance improvements originate from studies where multiple deep learning architectures and/or (pre)training schemes are evaluated on a
single dataset \citep{abousamra2022deep, saltz2018spatial, chen2020classification, bayramoglu2016deep, hameed2020breast}. This is helpful to identify the modeling choices that are beneficial for the given dataset, but it does not inform whether the modeling innovations would generalize across different datasets and/or classification tasks.

To address this gap, here we introduce ChampKit (Comprehensive Histopathology Assessment of Model Predictions toolKit). ChampKit enables systematic and reproducible comparisons of model architectures and transfer learning across several patch-based image classification datasets. ChampKit provides easy access to a large collection of models and networks from \verb|timm| \citep{rw2019timm}. ChampKit facilitates transfer learning/fine-tuning of those models for multiple patch-level classification tasks. The current implementation supports six diverse classification tasks across various cancers through publicly available datasets. ChampKit is extensible and allows integration with other custom models and datasets for benchmarking and comparison purposes. \textbf{ChampKit can greatly advance our ability to systematically identify the modeling innovations that generalize across patch-level histopathology classification tasks.}

\section{Related work}

\paragraph{Model repositories.} A large number of deep neural networks are available in popular model repositories \citep{NEURIPS2019_9015, tensorflow2015-whitepaper, huggingface, chollet2015keras, rw2019timm}. These make it easy to employ models pretrained on ImageNet for image classification or for transfer learning for other tasks. One such repository, \verb|timm| \citep{rw2019timm}, provides hundreds of image classification models implemented in PyTorch \citep{NEURIPS2019_9015} --- with many models pretrained on ImageNet-1K or -22k --- and training scripts for reproducibility. ResNet-based models are a popular choice for patch-level classification tasks in histopathology \citep{abousamra2022deep, kather2019deep, sharmay2021histotransfer, springenberg2022cnns, kassani2019classification}, while vision transformers have emerged as a candidate in recent studies \citep{zeid2021multiclass, chen2022gashis, springenberg2022cnns, deininger2022comparative}.

\paragraph{Comparative studies of models in histopathology.} \citet{laleh2022benchmarking} present benchmarks of many models for several multiple-instance-learning tasks, including patient-level survival prediction. These predictions are based on slide-level predictions, which are not directly comparable with patch-level classification tasks. \citet{sharmay2021histotransfer} evaluate transfer learning in nine CNNs on two datasets. The nine CNNs are limited to variations of ResNets \citep{he2016deep}, DenseNets \citep{huang2017densely}, and EfficientNets \citep{tan2019efficientnet}, and of the two datasets, one is not publicly available and the other is Camelyon \citep{bejnordi2017diagnostic}, which is included in our benchmarks (task 1). \citet{deininger2022comparative} compare a vision transformer (DeiT \citep{pmlr-v139-touvron21a}) to ResNet18 on two public datasets (a colorectal tissue classification dataset \cite{kather_jakob_nikolas_2018_1214456} and Camelyon). All of the models achieve accuracies greater than 0.998 area under the precision-recall curve on the colorectal cancer dataset, which suggests that it is not a meaningful dataset for model comparisons (We reproduced the near-perfect performance and therefore decided to exclude that datset from the current report). The other datasets in their study are not publicly available. \citet{kassani2019classification} evaluate ensembles of CNNs (i.e., VGG19 \citep{simonyan2014very}, MobileNet \citep{howard2017mobilenets}, and DenseNet \citep{huang2017densely}). All four datasets are broadly based on benign versus malignant tissue classification, and a code repository is not provided in the paper. \citet{springenberg2022cnns} provide an evaluation of several CNNs and vision transformers across five tasks, two of which are included in this study. Although their study is comprehensive, it does not include an evaluation of transfer learning, which is a common practice in histopathology analysis \citep{abousamra2022deep, kather2019deep, sharmay2021histotransfer, springenberg2022cnns, kassani2019classification, saltz2018spatial}.

In contrast to the studies above, we have curated a set of six diverse datasets for patch-level classification tasks that are publicly available. In addition, our benchmarking toolkit enables the evaluation of hundreds of different models, available in \verb|timm|, including models pretrained on ImageNet-1K and -22K \citep{ILSVRC15}. All of this is contained within a user friendly benchmarking toolkit that enables highly reproducible analyses, with extensions to custom models and datasets.

\section{ChampKit}

ChampKit is a one-stop-shop benchmarking toolkit that enables a systematic exploration of generalization performance of modeling choices and transfer learning across six patch-level histopathology classification tasks. ChampKit integrates the \verb|timm| \citep{rw2019timm} model repository to provide access to hundreds of (pretrained) deep learning models, enabling evaluation across different transfer learning schemes from ImageNet \citep{ILSVRC15} or from self-supervision on histopathology images \citep{ciga2022self, ciga2022github} or training the models from scratch.  The datasets for each task are curated from different studies \citep{kaczmarzyk2022dataset,kather_jakob_nikolas_2019_2530835,kather_jakob_nikolas_2019_2532612,veeling2018rotation,wei2021petri} and were selected based on diversity of the tasks and their importance for the biomedical community. Importantly, ChampKit provides insights into the modeling choices that are generalizable across patch-level classification tasks (not optimizing on a single dataset).

\paragraph{Workflow.} After one chooses a model and pretrained weights (if any), the model is systematically trained on all six tasks (one model for each task). ChampKit includes scripts to perform an end-to-end analysis: download the models, prepare the datasets, train the models, and evaluate them on held-out test data. Weights and Biases \cite{wandb} is used for logging experiment parameters and  visualizing results across multiple models and training configurations.

\paragraph{Custom models.} In addition to the popular (pretrained) computer vision models included in PyTorch, \verb|timm| provides access to a much larger selection of models that have been pretrained on different datasets (e.g. ImageNet-1K and ImageNet-22K, and self-supervised pretraining on ImageNet-22K). Nevertheless, it may be desirable to tweak an existing model or employ a custom model. To include a custom model to ChampKit, a PyTorch \citep{NEURIPS2019_9015} implementation of the model is required, as well as configuration options (i.e., input image size, location of pretrained weights if they exist, and mean and standard deviation of input normalization). The ChampKit code repository includes an example of adding custom pretrained weights for a ResNet18 model \citep{ciga2022github}.

\paragraph{Custom datasets.} Beyond the six benchmark datasets, ChampKit allows for integration with custom datasets that are framed as a single-task patch-level (binary or multi-class) classification. Custom datasets must be organized in an ImageNet-like directory structure, in which the top-level directory contains directories \verb|train|, \verb|val|, and \verb|test|, and each of those contains sub-directories of the different classes (e.g., \verb|tumor-positive|, \verb|tumor-negative|), where the corresponding patch-level images are located. Images can be of any size and will be resized during training and evaluation -- the size is configurable.

\section{Experiments}

Using ChampKit, we provide strong baselines for each benchmark task using two models commonly used in histopathology, namely ResNet18 and ResNet50 \citep{he2016deep}, and a hybrid vision transformer (R26-ViT) \citep{dosovitskiy2020image} that has shown promise at image classification in smaller data regimes, which is similar in spirit to the patch-level analysis, but has not been widely applied to histopathology. The scope of this study is to demonstrate the utility of ChampKit and provide benchmarking baselines on a variety of datasets and models. We do not seek to make claims of state-of-the-art in any task or provide practice recommendations. ChampKit is a powerful benchmarking toolkit that enables a systematic analyses of many models across diverse datasets and could be used for this purpose in future work.

All models included comparisons of transfer learning from image classification on ImageNet-1K \citep{ILSVRC15} and training from random weight initialization. ResNet18 comparisons also included transfer learning from a model trained using self-supervision on histopathology data \citep{ciga2022self, ciga2022github}. All models were implemented in \verb|timm| using PyTorch, and pretrained ImageNet weights were accessed via \verb|timm|, while the weights from self-supervised pre-training for ResNet18 were downloaded from \citep{ciga2022github}.

All models were trained systematically using similar training hyperparameters in order to make fair comparisons across models and datasets. Models were trained to minimize the cross-entropy loss using the AdamW optimizer \citep{loshchilov2017decoupled}. The learning rate was initially 1e-6 and warmed up to 1e-4 over the first three epochs, followed by a cosine schedule to a minimum value of 1e-6, set for 500 epochs. All models employed a dropout rate of 0.3,  DropPath \citep{huang2016deep} rate of 0.3, minibatch size of 84, label smoothing \citep{szegedy2016rethinking} of 0.1, and early stopping with a patience of 20 epochs based on validation loss. Automatic mixed precision was used via PyTorch's \citep{NEURIPS2019_9015} native implementation. Weights and Biases \citep{wandb} was employed to log experiment parameters and to track training and validation metrics.

In addition, data was augmented using the RandAugment strategy \citep{cubuk2020randaugment} (magnitude 9, standard deviation 0.5), random cropping (to at most 95\% of the image), and random erasing \citep{zhong2020random} (with a probability of 0.2 that a rectangle with an area between 1/50th and 1/3rd of the original image will be replaced with values sampled from a standard normal distribution). RGB channels were normalized with means (0.485, 0.456, 0.406) and standard deviations (0.229, 0.224, 0.225); these values are the means and standard deviations of the ImageNet training set. When using the R26-ViT or self-supervised ResNet18, images were normalized with means and standard deviations of (0.5, 0.5, 0.5) to match the normalization used in the pretraining. Images were resized to 224x224 with bicubic interpolation for R26-ViT and ResNet50, and bilinear interpolation for ResNet18.

Evaluation was done with model weights at 32-bit float precision based on early stopping. Area under the receiver operating characteristic curve (AUROC) and F1-score (threshold=0.5) were calculated (using \verb|torchmetrics| \citep{detlefsen2022torchmetrics}.
Each experiment was run on a single NVIDIA Quadro RTX 8000 GPU with 48GB of video memory.
A complete list of software versions can be found in the code repository.

\section{Benchmark tasks and results}

The benchmark datasets include six patch-based image classification tasks for: (1) tumor (versus no tumor) classification, (2) tumor-infiltrating lymphocyte detection, (3--5) microsatellite instability detection across different cancers and/or preparations, and (6) precancerous versus benign classification. ChampKit includes reproducible scripts to download all datasets, with the exception of the MHIST dataset \citep{wei2021petri}, which requires completing an online form (an automated email is then sent with a download URL). Details of each task are described alongside the results in the following subsections.

\subsection{Task 1: tumor (versus no tumor) classification}

\begin{table}[b!]
	\begin{minipage}{0.5\linewidth}
		\centering
		\includegraphics[width=1\textwidth]{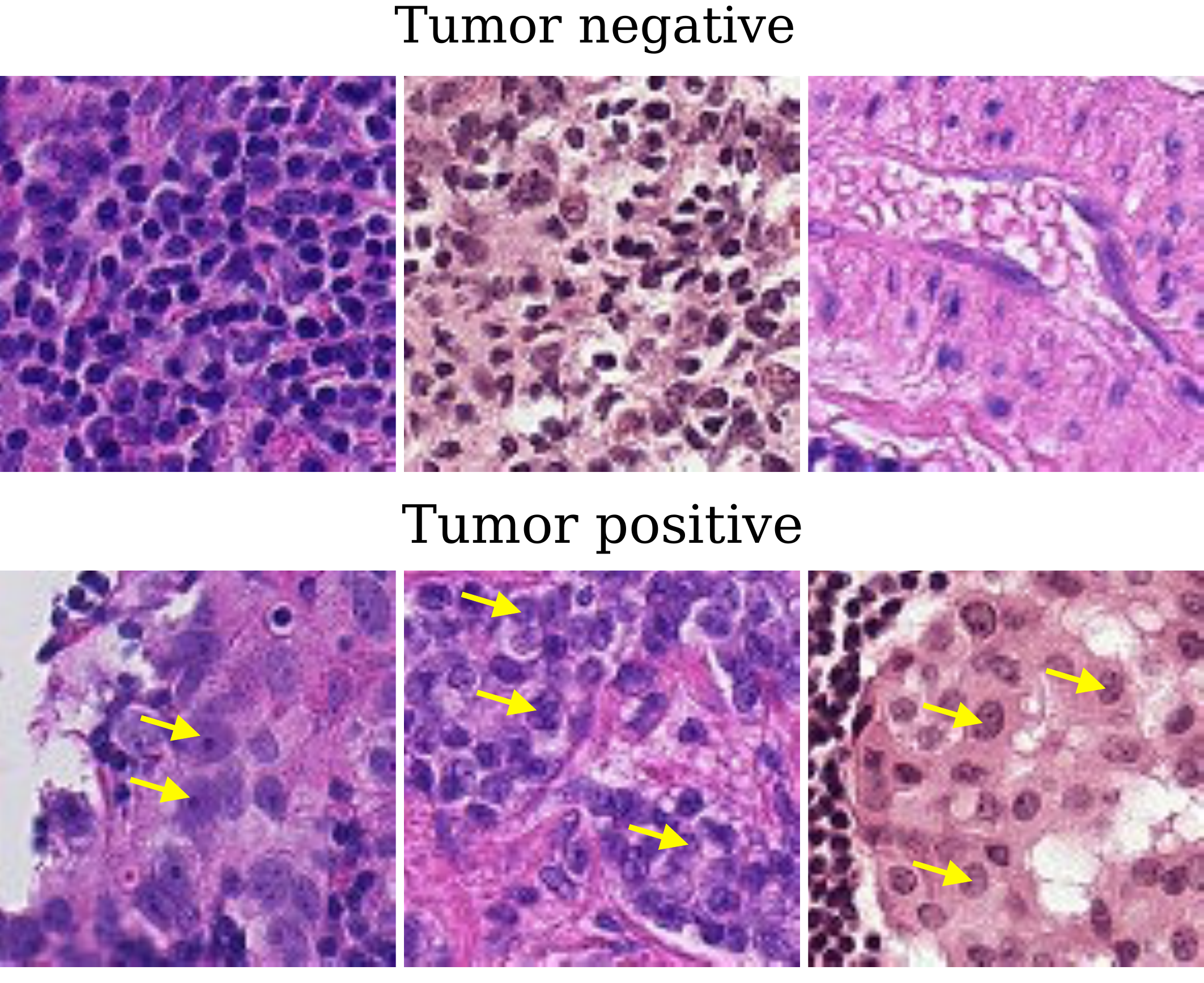}
		\vspace{1pt}
		\captionof{figure}{Sample tumor negative and tumor positive images from PatchCamelyon dataset \citep{veeling2018rotation}. The yellow arrows point to examples of tumor cells.}
		\label{figure-patchcam}
	\end{minipage}\hfill
	\begin{minipage}{0.48\linewidth}
        \centering
    		\resizebox{\textwidth}{!}{%
    	\begin{tabular}{llll}
        \toprule
        Model    & Pretraining & AUROC  & F1 score  \\
        \midrule
        R26-ViT  & None     & 0.951 & 0.845  \\
        R26-ViT  & ImageNet & \textbf{0.963} & \textbf{0.910}  \\
        \midrule
        ResNet50 & None     & 0.913 & 0.775  \\
        ResNet50 & ImageNet & \textbf{0.963} & \textbf{0.872}  \\
        \midrule
        ResNet18 & None     & 0.900 & 0.803  \\
        ResNet18 & ImageNet & 0.947 & 0.829  \\
        ResNet18 & SSL      & \textbf{0.956} & \textbf{0.858}  \\
        \bottomrule
        \end{tabular}
        }
        \vspace{1pt}
    	\caption{Results on task 1. Tumor (versus no tumor) classification in the PatchCamelyon dataset \citep{veeling2018rotation}.}
    	\label{table-patchcam}
	\end{minipage}
\end{table}

Detection of tumor cells is critical in clinical histopathology. Tumor cells can have varied appearances and can be challenging to detect. In particular, small nests of tumor cells (<100 cells) might be difficult to detect, and this is one case where automated deep learning algorithms can be highly useful. This is especially true in sentinel lymph node biopsies, which are performed to determine whether cells from the primary tumor have metastasized. False negatives are unacceptable in this situation, and so deep learning methods for this task must be rigorously evaluated. Deep learning methods for tumor detection can also improve tumor detection sensitivity and potentially reduce false negatives \citep{liu2019artificial}. We have included tumor detection as task 1 in our benchmark because of its clinical importance \citep{capbreastprotocol} and already wide-spread application in deep learning.

\paragraph{Dataset.}
The PatchCamelyon dataset \citep{veeling2018rotation} is a processed and curated version of the Camelyon16 dataset \citep{bejnordi2017diagnostic}, containing 327,680 tumor and non-tumor images at $96 \times 96$ pixels (10x magnification) from sentinel lymph node biopsies of breast cancer (Figure \ref{figure-patchcam}, Table \ref{dataset-sizes}  in Appendix).
An image is positively labeled if the center $32 \times 32$ pixel region contains at least one pixel of tumor.
The PatchCamelyon dataset is licensed under Creative Commons Zero v1.0 Universal and is anonymized.
 PatchCamelyon is available for download \cite{patchcamelyon} via Zenodo.

\paragraph{Results.}
For task 1, most models performed well according to AUROC and had more varied F1 scores (Table \ref{table-patchcam}). Since false negatives is an unwanted outcome for tumor classification, the F1 score is a more relevant metric. Interestingly, pretraining consistently led to higher performance than training from randomly initialized weights. For ResNet18, self-supervised pretraining on histopathology improved performance over ImageNet pretraining, consistent with findings in \citep{sharmay2021histotransfer}. Nevertheless, the R26-ViT architecture provides a strong inductive bias for this dataset (even without pretraining), performing as well as the best ResNet18 model with transfer learning; R26-ViT with pretraining yielded the best overall performance.

\subsection{Task 2: tumor-infiltrating lymphocyte detection}

\begin{table}[b]
	\begin{minipage}{0.5\linewidth}
		\centering
		\includegraphics[width=1\textwidth]{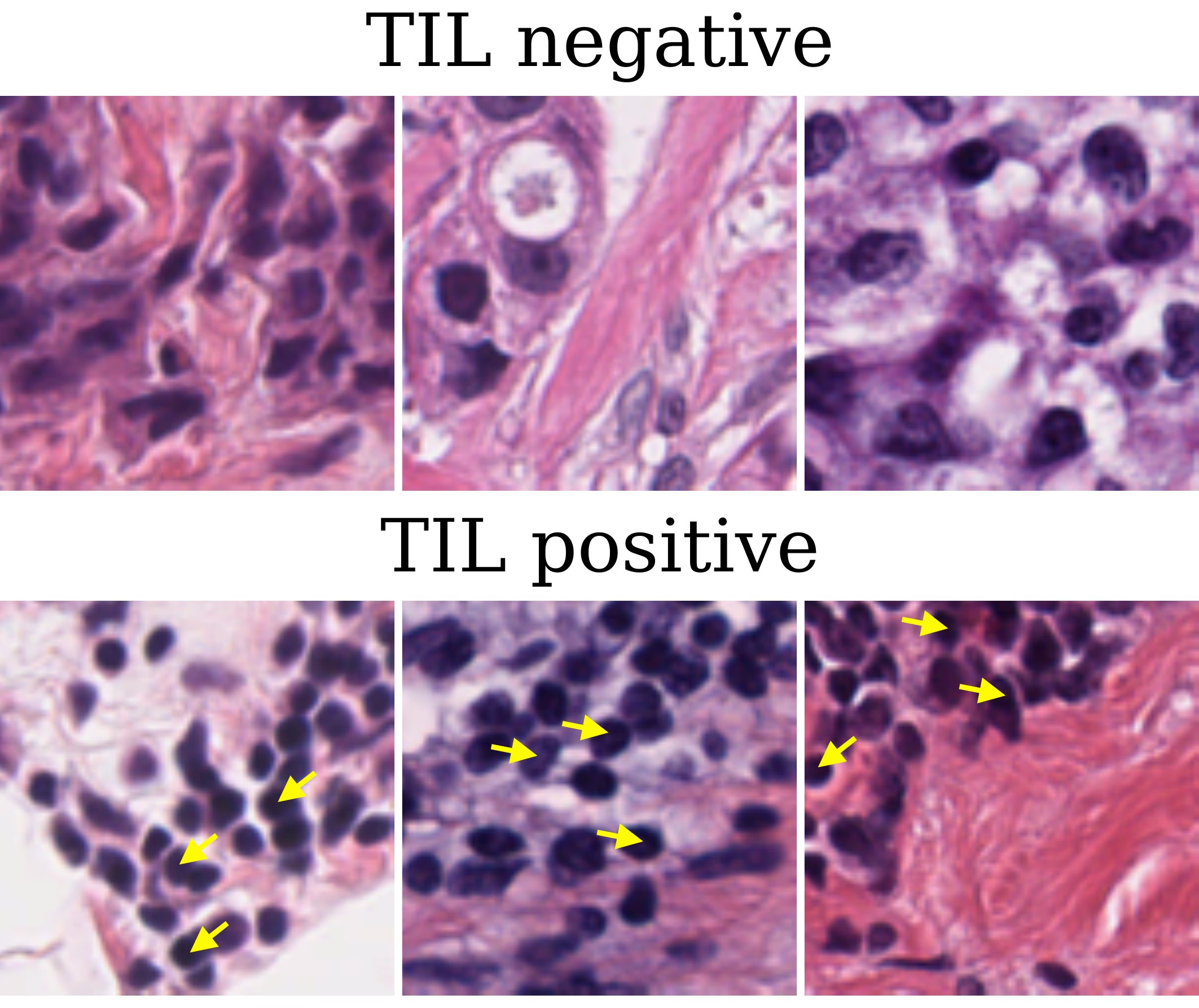}
		\vspace{1pt}
		\captionof{figure}{Sample images from TILs dataset \citep{kaczmarzyk2022dataset}. The yellow arrows point to examples of TILs.}
		\label{figure-tils}
	\end{minipage}\hfill
	\begin{minipage}{0.48\linewidth}
        \centering
    		\resizebox{\textwidth}{!}{%
    	\begin{tabular}{llll}
            \toprule
            Model    & Pretraining & AUROC  & F1 score  \\
            \midrule
            R26-ViT  & None     & \textbf{0.968} & \textbf{0.811}  \\
            R26-ViT  & ImageNet & 0.962 & 0.805  \\
            \midrule
            ResNet50 & None     & 0.968 & 0.797  \\
            ResNet50 & ImageNet & \textbf{0.973} & \textbf{0.837}  \\
            \midrule
            ResNet18 & None     & 0.975 & 0.835  \\
            ResNet18 & ImageNet & \textbf{0.983} & \textbf{0.863}  \\
            ResNet18 & SSL      & 0.981 & 0.855  \\
            \bottomrule
        \end{tabular}
        }
        \vspace{10pt}
    	\caption{Results on task 2 (TIL detection).}
    	\label{table-tils}
	\end{minipage}
\end{table}

Tumor-infiltrating lymphocytes (TILs) are clinically useful as prognostic biomarkers, related to the degree of immune response against a cancer. TIL quantification is important for predicting survival outcomes and guiding treatment decisions \citep{idos2020prognostic,paijens2021tumor,pages2018international,shaban2019novel}. TILs tend to be 8-12 $\mathrm{\mu m}$ in diameter with a dark, ovoid nucleus and scant cytoplasm \citep{robbins}. Despite the subtle qualitative differences of TILs across image patches, pathologists can identify TILs through visual inspection. However, in practice, they tend to characterize only a small number of microscopic fields of view \citep{capbreastprotocol}. More detailed prognostic patterns can be made by mapping TILs at a whole-slide-level \citep{fassler2022spatial}. Thus, it would be greatly beneficial to clinicians to identify the patches that contain TILs across histopathology slides \citep{salgado2015evaluation, saltz2018spatial, abousamra2022deep,pages2018international}. Deep learning has the potential to address major drawbacks of manual TIL scoring: inter-observer variability and the scalability of TIL detection. In response, there has been much interest in applying deep neural networks to this task \citep{saltz2018spatial, le2020utilizing, ZHANG2022102415, lu2020deep, abousamra2022deep,paijens2021tumor}. Thus, task 2 consists of pan-cancer TIL detection because of its tremendous clinical relevance and popularity in deep learning.

\paragraph{Dataset.}
Task 2 dataset consists of 304,097 TIL-positive and TIL-negative images from \citep{kaczmarzyk2022dataset}, a curated subset of the data presented in \citep{abousamra2022deep,saltz2018spatial} (Figure \ref{figure-tils}, Table \ref{dataset-sizes} in Appendix).
This dataset includes 23 different cancer types from The Cancer Genome Atlas (TCGA) \citep{hoadley2018cell}, representing a wide distribution of tissue types and stain differences. Patches are from formalin-fixed, paraffin-embedded (FFPE) whole slide images. Images are $100\times 100$ pixels at 0.5 $\mathrm{\mu m/pixel}$ and are positive if they contain at least two TILs. No stain normalization was applied to the images. The data is licensed under Creative Commons Attribution 4.0 International. Images are anonymized, and there is no overlap in TCGA participants across data splits. This dataset is available  for download via Zenodo \citep{kaczmarzyk2022dataset}.

\paragraph{Results.}
In general, all of the models do well at task 2, suggesting that there is strong predictive signal in the dataset (Table \ref{table-tils}). In contrast to task 1, the ResNet18 architecture is best overall on this TIL detection task, even without pretraining. This agrees with \citep{sharmay2021histotransfer}, which reports that smaller models tend to perform better on some histopathology tasks. Pretraining with ImageNet and self-supervised learning provide an additional performance gain. However, in contrast to the findings in \citep{sharmay2021histotransfer}, self-supervised pretraining did not outperform ImageNet pretraining. Interestingly, training from randomly initialized weights outperformed transfer learning for R26-ViT. This is consistent with the findings of \citet{kornblith2019better} and \citet{raghu2019transfusion} that ImageNet pretraining does not necessarily provide large performance gains. In addition, the scale of the features may play a role at the effectiveness of ImageNet-based transfer learning (i.e. task 2 images are magnified twice as much as those in task 1); further investigation is needed to better understand this interplay.

\subsection{Tasks 3--5: microsatellite instability detection}

Microsatellite instability (MSI) is an important prognostic clinical biomarker and has generated strong interest in recent years. MSI causes an abundance of DNA mutations and the formation of neoantigens, which activate the immune system, and causes changes in tissue morphology \citep{germano2017inactivation,robbins,alexander2001histopathological}. MSI is a useful clinical biomarker and is an indicator for PD-1/PD-L1 blocking therapies, like pembrolizumab \citep{casak2021fda, o2022pembrolizumab, luchini2019esmo, pietrantonio2021predictive, diaz2018keynote}. \citet{cercek2022pd1} recently found that their PD-1-blocking therapy led to remission in all 18 study participants. If a pathologist suspects an MSI phenotype, the standard of care is to conduct confirmatory molecular testing. Previously, \citet{kather2019deep} found that they could potentially avoid the time and cost of molecular testing by detecting MSI directly from histopathology. Many similar studies have been conducted \citep{echle2020clinical, yamashita2021deep, cao2020development, jenkins2007pathology, shia2003value, hyde2010histology, alam2022recent}, highlighting the importance of and excitement around MSI. We have included MSI detection in different cancer types and/or tissue preparations as tasks 3--5 because of the strong interest in predicting MSI from histopathology and the clinical relevance of MSI.

\begin{wrapfigure}{r}{0.53\textwidth}
  \begin{center}
    \includegraphics[width=0.53\textwidth]{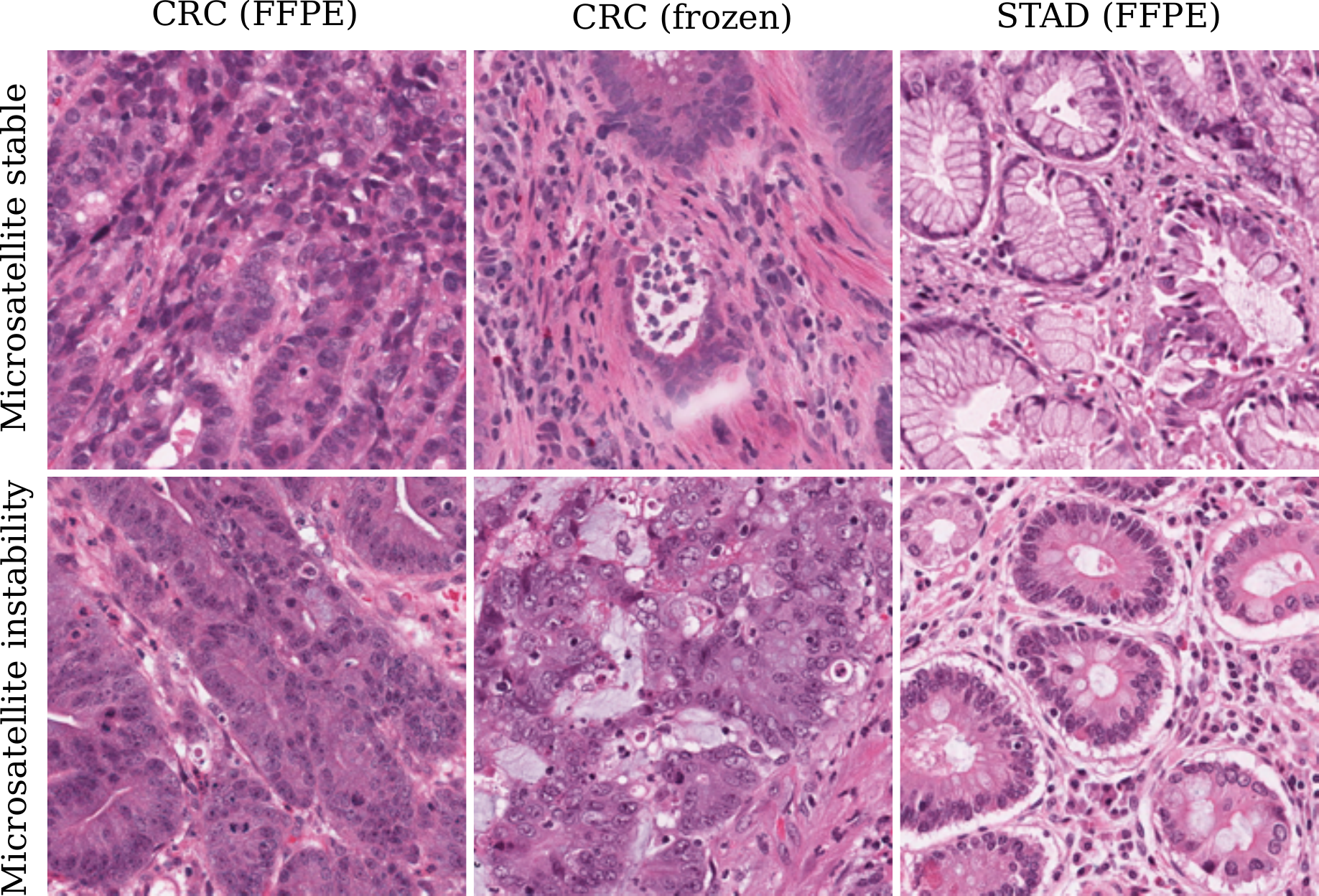}
  \end{center}
  \caption{Sample images from MSI datasets. Histologic features of MSI include poorly differentiated cells, signet ring cells, mucinous histology, cribriforming, and lymphocytic infiltrate \citep{alexander2001histopathological}. FFPE and frozen are two different types of tissue preparations.}
  \label{figure-msi}
\end{wrapfigure}

\paragraph{Dataset.}
MSI data was curated from \citep{kather2019deep}, which includes images from formalin-fixed, paraffin-embedded samples of colorectal carcinoma (CRC) and stomach adenocarcinoma (STAD) \citep{kather_jakob_nikolas_2019_2530835} and images from frozen samples of CRC \citep{kather_jakob_nikolas_2019_2532612}. All images are $224\times 224$ pixels at 0.5 $\mathrm{\mu m/pixel}$ (Figure \ref{figure-msi}, Table \ref{dataset-sizes} in Appendix). These datasets are publicly available and licensed under Creative Commons Attribution 4.0 International, and all images are anonymized. These datasets are available for immediate download via Zenodo (task 3 \citep{kather_jakob_nikolas_2019_2530835}, task 4 \citep{kather_jakob_nikolas_2019_2532612}, and task 5 \citep{kather_jakob_nikolas_2019_2530835}).

\paragraph{Results.}
Table \ref{table-msi} summarizes the results. ImageNet-pretrained ResNet50 consistently yielded the best performance across tasks. On the other hand, R26-ViT and ResNet18 yielded inconsistent results with pretraining. In some datasets, the best models were trained from random weight initialization. In other cases, ImageNet or self-supervised learning (SSL) pretraining was superior. The R26-ViT had competitive performance on task 3 (CRC --- FFPE) but yielded the worst performance on task 4 (CRC --- frozen) and task 5 (STAD). This suggests that the benefit of pretraining is task-specific.

Interestingly, \citet{kather2019deep} report patient-level AUROCs of 0.77 for CRC (FFPE), 0.84 for CRC (frozen), and 0.81 for STAD. We speculate that our AUROC values are lower because patch-level classification is a more difficult task than patient-level (i.e. whole-slide) classification.

The F1 scores are much lower, indicating poor precision and/or recall. Upon further inspection, there appears to be a wide test-validation gap; the AUROCs on the validation set were much higher (around 0.99). This suggests that the patch-level models are learning spurious correlations within individual patients (across training and validations sets) that do not generalize well to new patients in held-out test data. One strategy to combat this is to ensure that the validation set uses held-out patient data instead of a random split.

\begin{table}[t]
\centering
  \footnotesize
  \begin{tabular}{llllllll}
    \toprule
    & & \multicolumn{2}{c}{Task 3} & \multicolumn{2}{c}{Task 4} & \multicolumn{2}{c}{Task 5} \\
    & & \multicolumn{2}{c}{(CRC --- FFPE)} & \multicolumn{2}{c}{(CRC --- frozen)} & \multicolumn{2}{c}{(STAD --- FFPE)} \\
    \cmidrule(r){3-4}
    \cmidrule(r){5-6}
    \cmidrule(r){7-8}
    Model    & Pretraining & AUROC  & F1 & AUROC  & F1  & AUROC & F1  \\
    \midrule
    R26-ViT  & None     & \textbf{0.710} & \textbf{0.476} & 0.653 & \textbf{0.383}  & \textbf{0.660} & \textbf{0.429}  \\
    R26-ViT  & ImageNet & 0.657 & 0.408  & \textbf{0.704} & 0.367  & 0.658 & 0.396  \\
    \midrule
    ResNet50 & None     & 0.480 & 0.351  & 0.626 & 0.326  & 0.576 & 0.370  \\
    ResNet50 & ImageNet & \textbf{0.710} & \textbf{0.507}  & \textbf{0.736} & \textbf{0.475} & \textbf{0.747} & \textbf{0.513}  \\
    \midrule
    ResNet18 & None     & 0.693 & \textbf{0.468}  & 0.679 & \textbf{0.420}  & 0.685 & 0.425  \\
    ResNet18 & ImageNet & \textbf{0.708} & 0.454  & \textbf{0.688} & 0.397  & 0.705 & \textbf{0.479}  \\
    ResNet18 & SSL      & 0.681 & 0.456  & 0.679 & 0.396  & \textbf{0.717} & 0.465  \\
    \bottomrule
  \end{tabular}
        \vspace{10pt}
    \caption{Tasks 3--5 results. MSI detection in colorectal carconima (CRC) and stomach adenocarcinoma (STAD) with FFPE or frozen preparations.}
  \label{table-msi}
\end{table}

\subsection{Task 6: precancerous versus benign}

\begin{table}[b]
	\begin{minipage}{0.5\linewidth}
		\centering
		\includegraphics[width=1\textwidth]{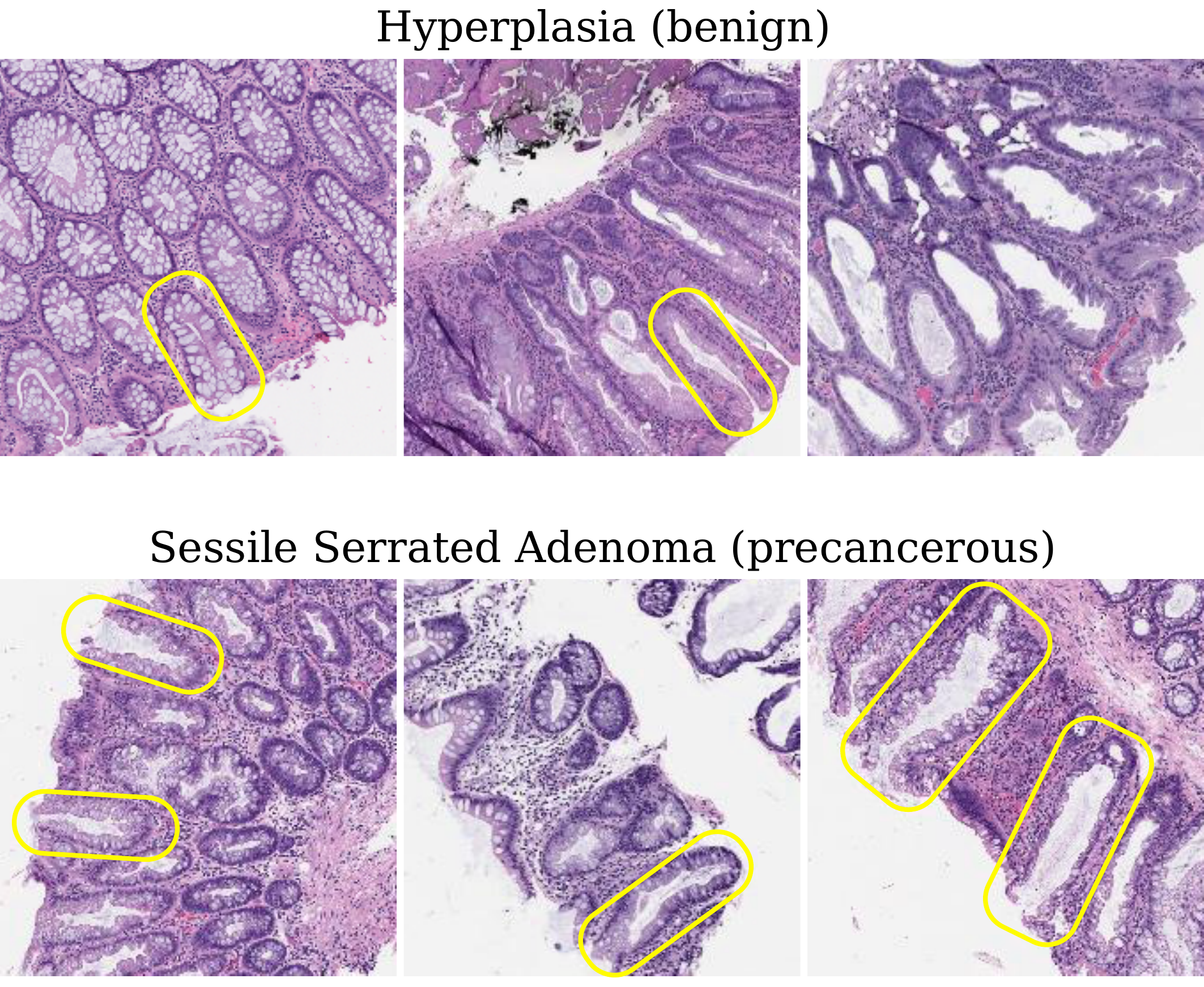}
		\vspace{1pt}
		\captionof{figure}{Sample images from MHIST dataset \citep{wei2021petri}.  Colonic crypts are outlined in yellow. Please note the difference in appearance between hyperplasia and adenoma.}
		\label{figure-mhist}
	\end{minipage}\hfill
	\begin{minipage}{0.48\linewidth}
        \centering
    		\resizebox{\textwidth}{!}{%
          \begin{tabular}{llll}
            \toprule
            Model    & Pretraining & AUROC  & F1 score  \\
            \midrule
            R26-ViT  & None     & 0.680  & 0.000  \\
            R26-ViT  & ImageNet & \textbf{0.882} & \textbf{0.784}  \\
            \midrule
            ResNet50 & None     & 0.672 & 0.210  \\
            ResNet50 & ImageNet & \textbf{0.884} & \textbf{0.741}  \\
            \midrule
            ResNet18 & None     & 0.665 & 0.282  \\
            ResNet18 & ImageNet & \textbf{0.919} & \textbf{0.804}  \\
            ResNet18 & SSL      & 0.911 & 0.768  \\
            \bottomrule
        \end{tabular}
        }
        \vspace{10pt}
    	\caption{Results on task 6 (precancerous versus benign).}
    	\label{table-mhist}
	\end{minipage}
\end{table}

Colonoscopies are an important screening test for colorectal carcinoma. Polyps are commonly found during the procedure \citep{obuch2015sessile}, and these polyps can be benign, precancerous, or cancerous. It is critical to correctly classify a benign polyp from one with cancerous potential because cancerous polyps might indicate need for additional treatment, but distinguishing between these remains challenging \citep{jaravaza2020hyperplastic}. False negatives are unacceptable in this task, and as such, there is significant interest in using deep learning to robustly detect precancerous polyps \citep{yoon2022colonoscopic, wei2020evaluation, korbar2017deep}. Due to the clinical importance of detecting precancerous colorectal polyps and the growing interest in applying deep learning to this problem, we elected to use the MHIST dataset \citep{wei2021petri} for task 6.

\paragraph{Dataset.} This dataset includes images of hyperplastic polyps and sessile serrated adenomas (Figure \ref{figure-mhist}). Hyperplasia is a benign overgrowth of cells, and an adenoma is a precancerous, low-grade disordered growth of cells. MHIST consists of 3,152 images colorectal polyps. The images were labeled as hyperplastic or adenomas by seven pathologists, and a binary classification is made by majority vote. All images are $224 \times 224$ pixels at 8x magnification and are deidentified. The MHIST dataset is the smallest dataset included in ChampKit (Table \ref{dataset-sizes} in Appendix), and this provides a useful test of how well different models and pretraining strategies cope with a small data regime.
To access the dataset, one must complete an online form. The user should then receive an automated email with a link to download the dataset. Once the dataset is downloaded, ChampKit can be used to prepare the dataset and train and evaluate models on the data.

\paragraph{Results.}
Unlike in the previous tasks, pretraining dramatically improves performance across all models (Table \ref{table-mhist}), consistent with the original MHIST publication \citep{wei2021petri}. Resnet18 performed best overall, consistent with \citep{sharmay2021histotransfer}. It was difficult to train models from randomly initialized weights, and the low AUROC and F1 scores reflect this. The R26-ViT in particular achieved an F1 score of 0.00, suggesting that it is especially challenging to train an R26-ViT from scratch on small histopathology datasets. We speculate that pretraining was especially important here  because of the small dataset size. Pretraining might provide useful initializations for other small datasets.

\section{Conclusion}

Here we introduce ChampKit, a reproducible benchmarking toolkit for patch-based image classification in histopathology. We use ChampKit to provide baseline results for multiple models on six histopathology datasets. We found that transfer learning can improve classification performance, but this is not consistent across tasks. It remains unclear whether the scale of the histopathology features (i.e., magnification) plays a role in being amenable to transfer learning based on models pretrained on natural images. ChampKit enables the systematic evaluation of transfer learning on patch-based image classification, and we hope that it will greatly advance the knowledge of transfer learning and modeling innovations in histopathology.

In this study, the baseline comparisons were limited to three networks with different pretrained weights. Thus, strong claims of modeling choices cannot be drawn from such a small scale study. Nevertheless, ChampKit uses \verb|timm| to enable easy access to hundreds of other models with different pretrained weights that were not explored here. Hence, it should enable more systematic studies to investigate how model choices, such as small architectures vs large architectures, CNNs vs transformers, transfer learning with pretraining on ImageNet vs self-supervised pretraining vs random initialization, or other modeling choices. In addition, one training run is shown per model here. While these runs are fully reproducible when using the same random number seed, the distribution of training results using multiple seeds can help evaluate the robustness of model performance. Moreover, other patch-level histopathology tasks, including tumor subtype and grade classification, should also be added to expand the benchmark to other popular deep learning applications.

In summary, we hope that ChampKit accelerates research in deep learning and histopathology towards a future in clinical practice.

\begin{ack}

We gratefully acknowledge support from National Cancer Institute grants U24CA215109 and UH3CA225021.
PKK was supported in part by funding from the Simons Center for Quantitative Biology at Cold Spring Harbor Laboratory. JRK was also supported by National Institutes of Health grant T32GM008444 (NIGMS). JRK would also like to acknowledge the support of the Medical Scientist Training Program at Stony Brook University. The results shown here are in part based upon data generated by the TCGA Research Network: https://www.cancer.gov/tcga. We thank Shushan Toneyan for going through the source code and reproducing parts of this manuscript. Finally, we would like to thank Satrajit S. Ghosh for help in naming this project.
\end{ack}

\bibliography{main}

\appendix

\section{Appendix}

\begin{table}[ht]
  \caption{Dataset sizes.}
  \label{dataset-sizes}
  \centering
  \begin{tabular}{lllllll}
    \toprule
    Dataset & \multicolumn{2}{c}{Train} & \multicolumn{2}{c}{Val} & \multicolumn{2}{c}{Test} \\
    \cmidrule(r){2-3}
    \cmidrule(r){4-5}
    \cmidrule(r){6-7}
                        & Positive  & Negative  & Positive  & Negative  & Positive  & Negative  \\
    \midrule
    PatchCamelyon       & 131,072   & 131,072   & 16,369    & 16,399    & 16,377    & 16,391    \\
    TILs                & 39,206    & 170,015   & 5,203     & 33,398    & 10,501    & 45,774    \\
    MSI STAD            & 40,228    & 40,228    & 10,057    & 10,057    & 27,904    & 90,104    \\
    MSI CRC (FFPE)      & 37,363    & 37,363    & 9,341     & 9,341     & 28,335    & 70,569    \\
    MSI CRC (frozen)    & 24,357    & 24,358    & 6,090     & 6,089     & 17,675    & 60,574    \\
    MHIST               & 493       & 1,247     & 137       & 298       & 360       & 617       \\
    \bottomrule
  \end{tabular}
\end{table}

\end{document}